\pdfoutput=1
\documentclass[longbibliography,osajnl,preprint,superscriptaddress,12pt]{revtex4-1} 
\usepackage{amsmath,amssymb,graphicx}
\usepackage[utf8]{inputenc}
\begin{document}

\title{Shot noise reduced terahertz detection\\
via spectrally post-filtered electro-optic sampling}

\author{Michael Porer}
\author{Jean-Michel M\'{e}nard}
\author{Rupert Huber}

\affiliation{
Department of Physics, University of Regensburg, 93053 Regensburg, Germany\\
Corresponding author: michael.porer@gmail.com
}

\begin{abstract}
In ultrabroadband terahertz electro-optic sampling, spectral filtering of the gate pulse can strongly reduce the quantum noise while the signal level is only weakly affected. The concept is tested for phase-matched electro-optic detection of field transients centered at $45\,\mathrm{THz}$ with 12-fs near-infrared gate pulses in AgGaS$_2$. Our new approach increases the experimental signal-to-noise ratio by a factor of 3 compared to standard electro-optic sampling. Under certain conditions an improvement factor larger than 5 is predicted by our theoretical analysis. 
\end{abstract}

\maketitle  

One of the key achievements of terahertz (THz) photonics is the possibility to detect free-space electromagnetic radiation with respect to the absolute phase and amplitude of its oscillating carrier wave  \cite{Ferguson2002,Tonouchi2007}.
Among all technological implementations of field-sensitive detection \cite{Smith1988, Dai2006}, electro-optic sampling (EOS) \cite{Wu1995,Liu2004,Kuebler2004} stands out due to its excellent sensitivity. 
This technique has been broadly used in THz time-domain spectroscopy (TDS) \cite{Ulbricht2011}, sensing and imaging \cite{Tonouchi2007,Moon2012,Blanchard2013}. Furthermore electro-optic detection has facilitated ultrafast pump-probe studies that resonantly access important low-energy dynamics of condensed matter throughout the entire far- and mid-infrared spectral range \cite{Ulbricht2011,Kampfrath2013}. Steady technological progress has pushed the frequency bandwidth accessible to EOS close to the near-infrared (NIR) \cite{Sell2008,Matsubara2012}. For sufficiently strong THz fields, EOS readily allows for recording a complete THz waveform with a single laser shot \cite{Shan2000}. Further increase of the sensitivity may ultimately enable a novel research field of THz quantum optics including single-shot sampling of few-photon squeezed THz pulses or THz photon bunches emitted from the quantum vacuum \cite{Ciuti2005,Gunter2009,Porer2012}. Currently, the ultimate limit of detector sensitivity is set by the quantum granularity, i.e. the shot noise, of the gate laser pulses used for EOS \cite{Kuebler2004}. Recent efforts to optimize the detector performance have aimed at improved detection electronics \cite{Darmo2011}, but have not diminished the shot noise itself. Here, we introduce a novel method to reduce the shot noise in EOS. Our approach can lower the EOS acquisition time by more than one order of magnitude.
\par 

\begin{figure}[t]
\centerline{\hspace{0cm}\includegraphics[width=8.1cm]{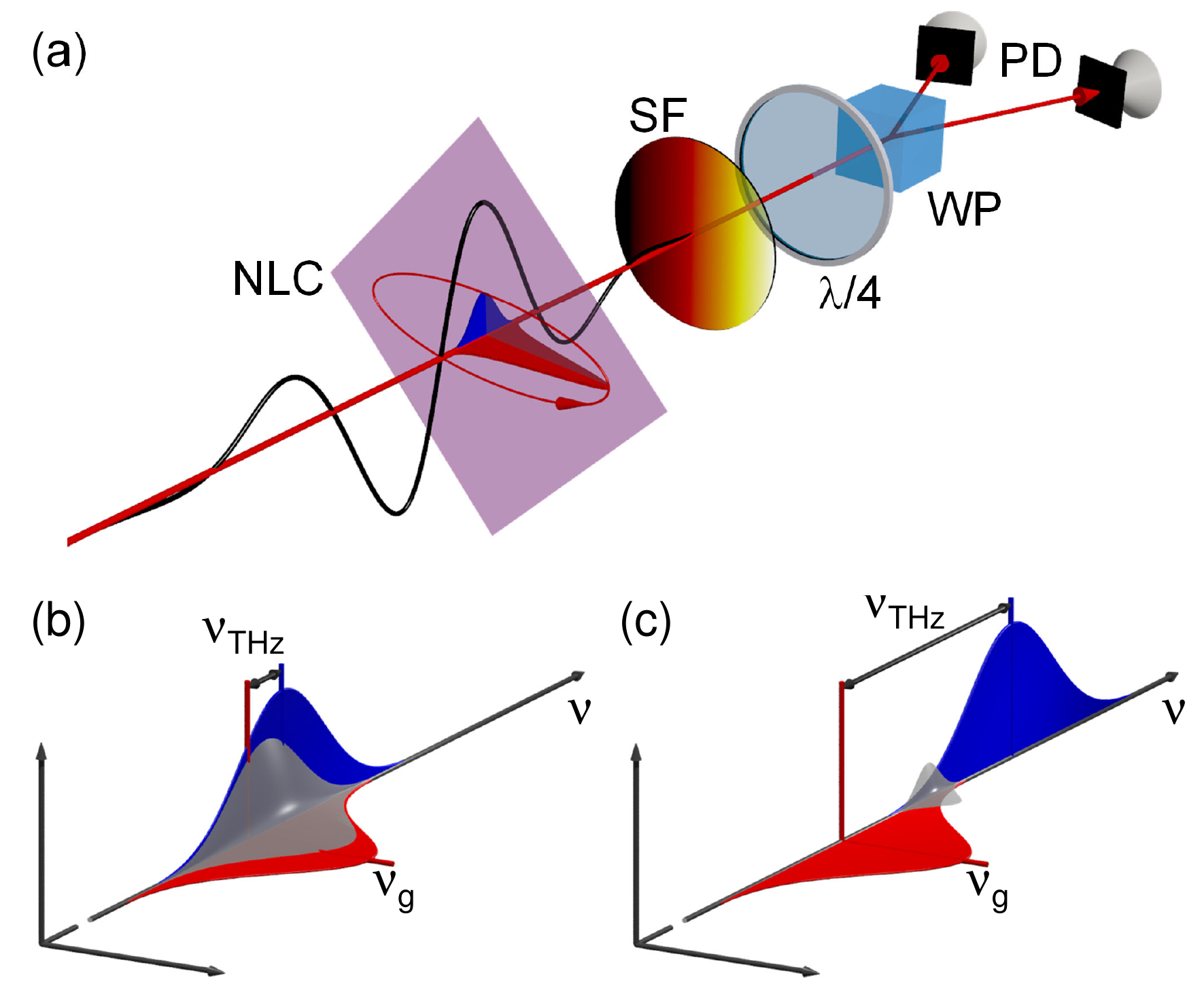}}\vspace{0mm}
\caption{(a) EOS setup consisting of a nonlinear crystal (NLC) and an ellipsometer ($\lambda/4$ plate; Wollaston prism (WP); photodiodes (PD)). Black waveform: THz field transient. The nonlinear optical interaction of the linearly polarized incident gate pulse (red pulse) with the THz wave generates new frequency components at perpedicular polarization (blue pulse) which induces an elliptical polarization of the gate (indicated by the red ellipse). For frequency post-filtered EOS, a spectral filter (SF) is inserted behind the NLC. (b,c) Schematic spectra of the incident gate pulse (red) (centered at $\nu_\mathrm{g}$) and of the thereto orthogonally polarized sum frequency photons (blue) when the THz frequency  $\nu_\mathrm{THz}$ is small (b) and comparable (c) to the bandwidth of the gate spectrum $\delta_\mathrm{g}$. The EO signal is located at the overlap of both spectra (gray shaded area). 
\label{setup}} 
\end{figure}

In EOS, a phase-stable THz transient is focused into a nonlinear optical crystal (NLC), where it co-propagates with an ultrashort optical gate  (Fig. \ref{setup}(a)) \cite{Gallot1999}. In a simplified picture, the THz electric field induces a quasi-instantaneous birefringence via the Pockels effect, causing a phase retardation $\Delta \varphi$ between the linear polarization components of the gate pulse. The interaction length between THz and gate pulses can be maximized by phase matching \cite{Liu2004,Kuebler2004}. $\Delta \varphi$ is read out with an ellipsometer consisting of a quarter wave plate and a Wollaston prism (WP). The polarization optics split the gate power equally between two identical photodiodes (PD) as long as no THz field is applied, while a THz-induced phase retardation $\Delta \varphi$ causes an imbalance $S=I_\mathrm{a}-I_\mathrm{b}$ of the photocurrents ($I_\mathrm{a},I_\mathrm{b}$) recorded in the diode pair. Repeating this experiment as a function of the \mbox{delay time $t$} between the THz and the gate pulse yields a differential signal $S(t)$ that is directly proportional to the time trace of the THz electric field $E_\mathrm{THz}(t)$. 
\par

Balanced differential detection largely eliminates technical noise, such as excess power fluctuations of the gate pulses, since it typically affects the photocurrents in both diodes equally. In state-of-the-art EOS, technical noise can be routinely suppressed to a level where the shot noise of the gate pulses becomes visible: The number of incident light quanta follows a non-deterministic quantum distribution on each of the two photodiodes and can, hence, not be fully balanced. For a train of coherent gate pulses with an average power $P_\mathrm{g}$ the Poissonian statistics generates shot noise $\Delta_\mathrm{S}$ that scales with $\sqrt{P_\mathrm{g}}$ \cite{Milonni2010}. Since the electro-optic (EO) signal itself depends linearly on $P_\mathrm{g}$ \cite{Gallot1999}, the maximum signal-to-noise ratio (SNR) should, in principle, rise with $\sqrt{P_\mathrm{g}}$. In practice, $P_\mathrm{g}$ is often limited by available laser power or undesired higher-order non-linear processes in the NLC, such as two-photon absorption. Furthermore, eliminating the excess noise sufficiently to reach the shot noise level is more challenging for high $P_\mathrm{g}$.

\par

We will now show that even for shot noise limited detection the SNR can be further enhanced in commonly used EOS configurations. In order to explain our idea, we describe EOS in the more rigorous picture of frequency mixing between the gate and the THz photons \cite{Gallot1999}. Depending on the interaction geometry, the $\chi^{(2)}$ susceptibility of the NLC can give rise to either sum or difference frequency generation between the gate and the THz pulse. These nonlinear interactions create a phase-coherent replica of the broadband gate spectrum which is up- or down-shifted by the THz frequency, respectively (Fig. \ref{setup}(b)). The newly generated photons are polarized perpendicularly to the incident gate light. Interference of converted and fundamental photons yields a modified polarization state in the frequency region where both spectra overlap. When the resulting light is analyzed in a base rotated by $\pi/4$ with respect to the incident gate polarization, the phase difference between the orthogonal polarization components ($\Delta \varphi$) scales linearly and sign-sensitively with the THz electric field. Note that the EO signal $S(t)$ is solely generated by photons from the spectral overlap region. When the frequency of the THz photons  $\nu_\mathrm{THz}$ is small compared to the bandwidth  $\delta_\mathrm{g}$ of the gate spectrum (FWHM of amplitude), the interference occurs over a large portion of the gate spectrum and the Pockels effect picture remains valid (Fig. \ref{setup}(b)). If $\nu_\mathrm{THz}$ is comparable to $\delta_\mathrm{g}$, the overlap region containing the EO signal is located only at the wings of the gate spectrum, reducing the signal level (Fig.  \ref{setup}(c)). Still all gate photons contribute equally to the shot noise. In this situation, a spectral filter (SF) inserted into the gate beam after the NLC (see Fig. \ref{setup}(a)) can be used to select photons that contribute to the EO signal and block those which only contribute to the noise. 
\par

\begin{figure}[t]
\centerline{\includegraphics[width=8.3cm]{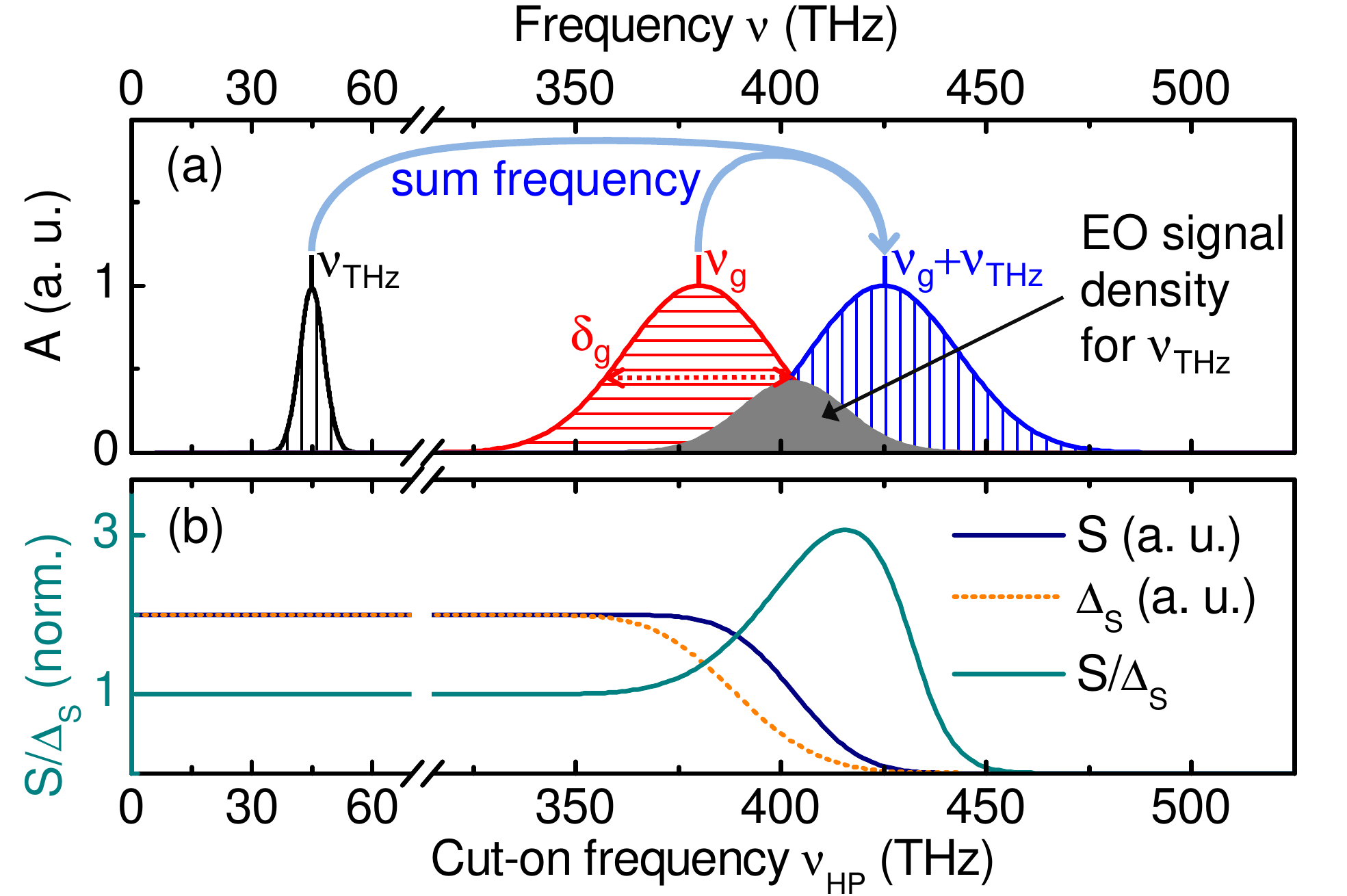}}\vspace{-2mm}
\caption{EOS in the case of SFG and $\nu_\mathrm{THz} \approx \delta_\mathrm{g}$. (a) Solid curves: Schematic amplitude spectra of the THz pulses (black), the incident NIR gate pulses (red) and the sum frequency replica of the gate spectrum (blue) for a given $\nu_\mathrm{THz}$. The hatching direction indicates the polarization direction in typical phase-matching geometries. Interference of the gate spectrum and its replica in the region of their spectral overlap (shaded area) yields elliptical polarization necessary for a field-sensitive EO signal. (b) Spectrally integrated EO signal for $\nu_\mathrm{THz}$ (blue curve) and shot noise level (orange dotted curve) as a function of the cut-on frequency of the high-pass SF, as calculated with the model described in the text. Cyan line: Normalized ratio between signal and shot noise. \label{frequencymixing}} 
\end{figure}

In order to quantify how spectral filtering improves the SNR, we analyze EOS assuming perfect phase matching for sum frequency generation (SFG), a frequency independent susceptibiltiy $\chi^{(2)}$ and bandwidth limited pulse durations. Figure \ref{frequencymixing}(a) schematically depicts typical spectra (solid lines) often seen in multi-THz spectroscopy based on Ti:sapphire lasers \cite{Kuebler2004,Liu2004}. The red line shows the incident gate spectrum $A(\nu)$ centered at a frequency of $\nu=\nu_\mathrm{g}$. The sum frequency spectrum $A(\nu-\nu_\mathrm{THz})$ resulting from a given THz component $\nu_\mathrm{THz}$ and the THz spectrum $A_\mathrm{THz}(\nu)$ are shown in blue and black, respectively. Under the above assumptions, the spectral density of the EO signal (gray shaded area) is proportional to $\nu A(\nu) A(\nu-\nu_\mathrm{THz})$, for a given frequency $\nu_\mathrm{THz}$ \cite{Gallot1999}. Integrating over $\nu$ yields the EO response as a function of $\nu_\mathrm{THz}$. In a conventional EOS setup, the photodiodes perform the spectral integration. If the gate pulses are post-filtered with a high-pass (HP) the lower integration boundary is set by the cut-on frequency $\nu_\mathrm{HP}$. In summary, the spectral signal amplitude   depends on the interacting pulses as follows:

\begin{align}
S \propto A_\mathrm{THz}(\nu_\mathrm{THz})\int^{\infty}_{\nu_{\mathrm{HP}}}\!\nu  A(\nu)A(\nu\!-\!\nu_\mathrm{THz})\mathrm{d}\nu\nonumber
\end{align}

The shot noise recorded with the SF amounts to:  

\begin{align}
\Delta_\mathrm{S}\propto\sqrt{\int\limits^{\infty}_{\nu_{\mathrm{HP}}}\!\!\left( (1\!\!-\!\!\eta)\left|A(\nu)\right|^2+\eta\left|A(\nu\!-\nu\!_\mathrm{THz})\right|^2\right)\mathrm{d}\nu}\nonumber
\end{align}

Here we explicitly take into account that for large quantum efficiencies $\eta$ of the sum frequency process, a large number of incident gate photons may be converted from frequencies below to above $\nu_\mathrm{HP}$ and pass the filter. Figure \ref{frequencymixing}(b) compares $S$ (blue curve) and  $\Delta_\mathrm{S}$ (orange dots) as a function of $\nu_\mathrm{HP}$ for $\nu_\mathrm{THz}=45\,\mathrm{THz}$, $\eta=1\times10^{-4}$ and the gate spectrum of Fig. \ref{frequencymixing}(a). $S$ remains unaffected for $\nu_\mathrm{HP}\ll\nu_\mathrm{g}$, but decreases as $\nu_\mathrm{HP}$ shifts through the overlap region between the fundamental and sum frequency spectra. In contrast, the noise drops faster with increasing $\nu_\mathrm{HP}$. A clear reduction is already seen when $\nu_\mathrm{HP}$ reaches the low-frequency wing of the incident spectrum, i.e. below the overlap region with the sum frequency spectrum. This leads to an improvement of the SNR (cyan curve), which exhibits a well-defined maximum before it drops towards zero. If no other noise source becomes dominant for a lower overall photon fluence, this peak identifies the cut-on frequency for maximum SNR.

\begin{figure}[t]
\centerline{\includegraphics[width=8.5cm]{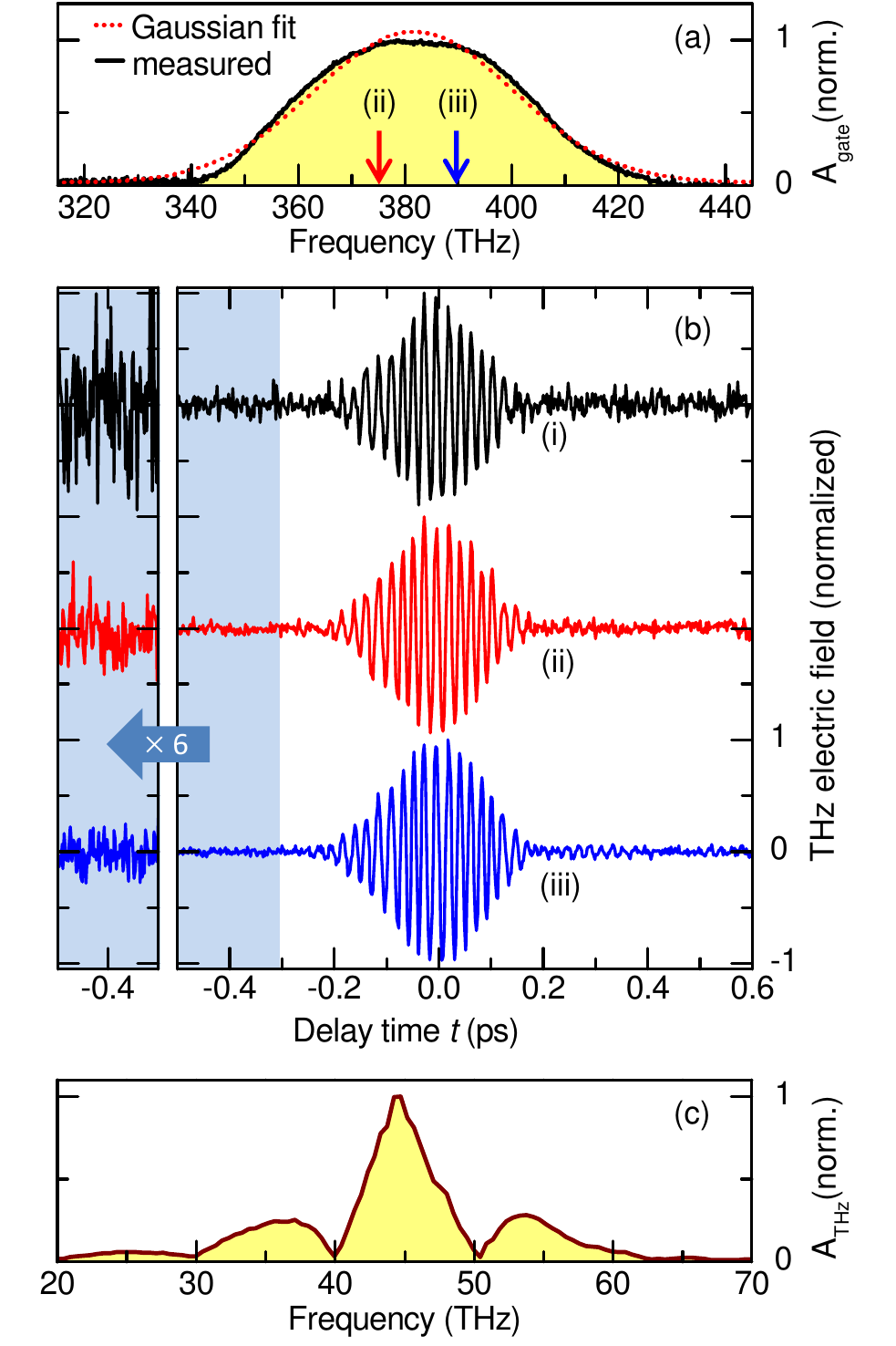}\vspace{-1mm}}
\caption{(a) Amplitude spectrum of the incident NIR laser pulses used for OR and EOS. Black curve: experimental data. Red dotted curve: Gaussian fit. Arrows at (ii) and (iii) indicate the cut-on frequencies of the high-pass filters used in the experiment. (b) Multi-terahertz field transient recorded with phase-matched EOS in AGS employing (i) no spectral filtering or (ii) high-pass filtering with a cut-on frequency of $375\,\mathrm{THz}$, and (iii) $390\,\mathrm{THz}$, respectively. For better visibility of the noise floor, a very weak THz signal is studied. The panel on the left magnifies the amplitude in the shaded time window by a factor of 6. (c) Linear amplitude spectrum of the unattenuated THz transients. \label{transients}} \vspace{-3mm}
\end{figure}

\par
To test this idea experimentally, we generate THz transients by phase-matched optical rectification (OR) of \mbox{12-fs} NIR pulses derived from a Ti:sapphire amplifier system \cite{Huber2003} ($f_\mathrm{rep}=800\,\mathrm{kHz}$, $E_\mathrm{pulse}=0.1\,\mathrm{\mu J}$) in a AgGaS$_2$ (AGS) crystal ($\phi=45^\circ$, $\theta=57^\circ$, thickness: $200\,\mathrm{\mu m}$) \cite{Fan1984, Bonvalet1995, Rotermund2000}. The phase-locked THz waveform is electro-optically detected in a second AGS crystal with the same orientation and a thickness of $100\,\mathrm{\mu m}$ (Fig. \ref{transients}(b)). In order to implement phase-matched SFG, the polarization of the incident THz and gate pulse is set to ordinary and extraordinary, respectively. The bandwidth and center frequency of the bandwidth-limited gate pulses \mbox{(Fig. \ref{transients}(b))} are \mbox{$\delta_\mathrm{g} = 42\,\mathrm{THz}$} and $\nu_\mathrm{g} = 380\,\mathrm{THz}$ \mbox{(Fig. \ref{transients}(a))}. 
A lock-in amplifier is used for electronic data readout and allows us to detect a THz induced imbalance on the photocurrents close to the shot noise which corresponds to \mbox{$\Delta_\mathrm{S}=(I_\mathrm{a}+I_\mathrm{b})\times9\times 10^{-9}\,\mathrm{Hz^{-1/2}}$}. In order to visualize the noise floor, we deliberately attenuate the THz power to an estimated number of THz photons of $\sim 10^6$ per pulse and choose a short lock-in time constant of $100\,\mathrm{\mu s}$. This way, we record the same few-cycle multi-THz waveform with three different choices of $\nu_\mathrm{HP}$ (curves (i)-(iii) in Fig. \ref{transients}(b)). Figure \ref{transients}(c) depicts the amplitude spectrum of the transient centered at $\nu_\mathrm{THz}=45\,\mathrm{THz}$. As seen from a comparison of the two curves labelled (i) and (ii), in Fig. \ref{transients}(b), the SNR is dramatically improved by optical filtering. Increasing $\nu_\mathrm{HP}$ from $375\,\mathrm{THz}$ (curve (ii)) to $390\,\mathrm{THz}$ (curve (iii)) yields further improvement. Quantitatively, the SNR increases by a factor of 2.9 for $\nu_\mathrm{HP} = 390\,\mathrm{THz}$ as compared to conventional unfiltered EOS. With a lock-in integration time of $1\,\mathrm{s}$ the sensitivity achieved here suffices to detect the coherent signal of a train of THz transients containing less than one photon per pulse. For the experimental parameters ($\nu_\mathrm{g} = 380\,\mathrm{THz}$, $\nu_\mathrm{THz}=45\,\mathrm{THz}$, $\delta_\mathrm{g} = 42\,\mathrm{THz}$, $\eta=1\times10^{-4}$, $\nu_\mathrm{HP} = 390\,\mathrm{THz}$), our theory predicts an improvement of the SNR by a factor of $1.9$. The experimental gain in SNR thus even exceeds the theoretical value. We attribute this fact to the suppression of a remaining fraction of technical noise carried by the gate light. Without SF the actual noise on the acquired transient is a factor of $\sim 1.5$ above the calculated shot noise level. Since the technical noise scales linearly with the power impinging on the photodiodes, reducing the overall photon fluence lowers the influence of technical noise as well. Accounting for both effects, our overall improvement by spectral filtering is perfectly explained.
\par

\begin{figure}[t]
\vspace{-1mm}\hspace{-3mm}\centerline{\includegraphics[width=8.4cm]{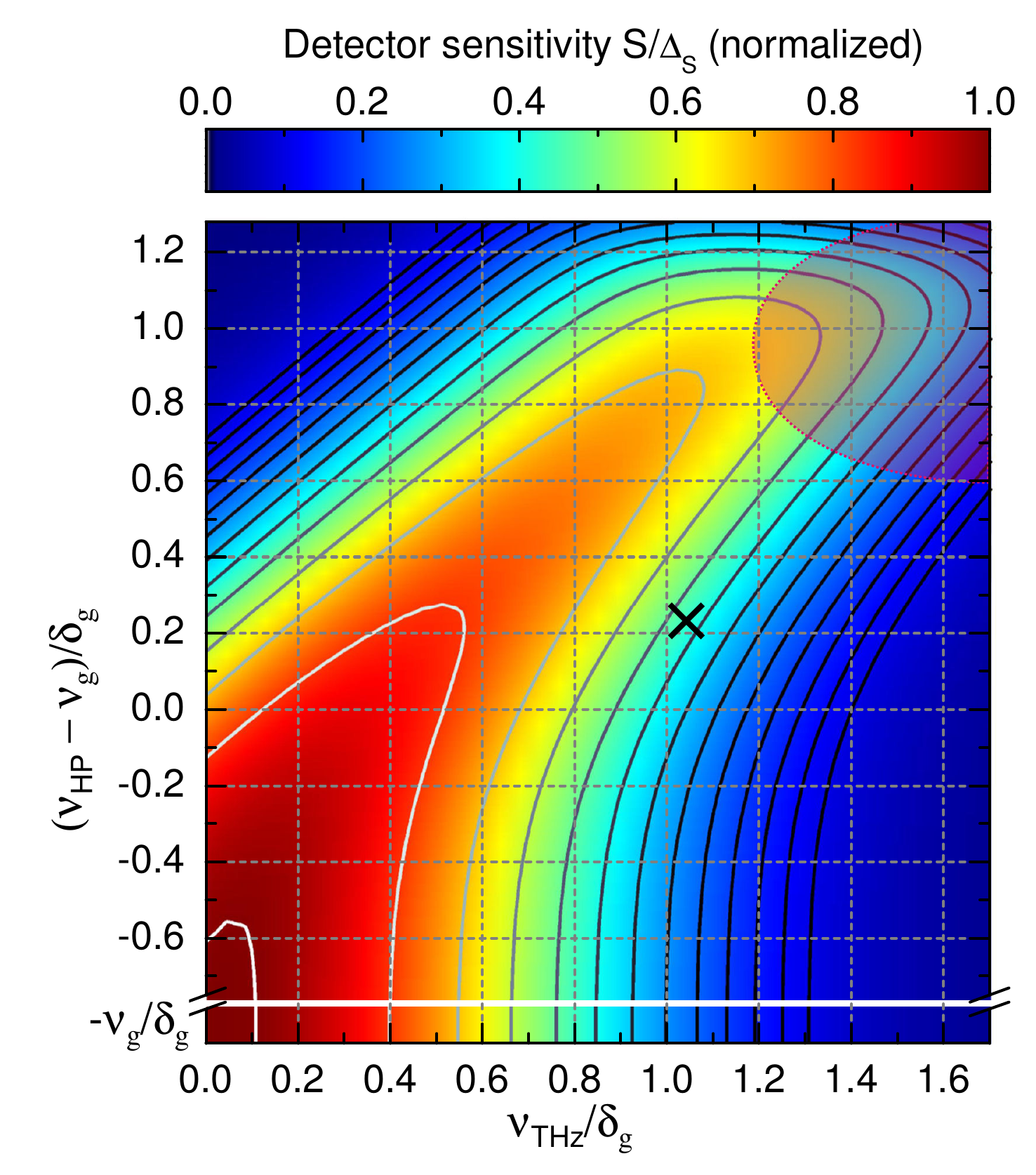}}\vspace{-4mm}
\caption{2D map of the calculated sensitivity $S/\Delta_\mathrm{S}$ of spectrally post-filtered EOS for the SFG process discussed in the text. The scale is normalized to the sensitivity obtained for a static electric field without SF. The abscissa corresponds to the THz frequency $\nu_\mathrm{THz}$ normalized to the gate bandwidth $\delta_\mathrm{g}$ (FWHM of amplitude spectrum). The ordinate represents the cut-on frequency of the HP filter, $\nu_\mathrm{HP}$, relative to the center frequency of the gate pulse, $\nu_\mathrm{g}$, normalized to $\delta_\mathrm{g}$. The region below the vertical scale break shows the sensitivity without SF. Contour lines are spaced by a factor of 1.2. The model holds for $\delta_\mathrm{g}\ll\nu_\mathrm{g}$ and $\eta=1\times10^{-4}$. The coordinate corresponding to our experiment with $\nu_\mathrm{HP} = 390\,\mathrm{THz}$ is indicated with a black cross. An increase of $S/\Delta_\mathrm{S}$ by more than a factor of $5$ can be obtained for configurations located in the purple shaded region. \label{sensitivitymap}} 
\end{figure}

Even larger improvement factors due to shot noise reduction alone may be expected for other experimental parameters. For the assumptions made above, Fig.  \ref{sensitivitymap} summarizes the theoretical gain in the SNR as a function of the relative THz frequency $\nu_\mathrm{THz}/\delta_\mathrm{g}$ and the relative difference $(\nu_\mathrm{HP} - \nu_\mathrm{g})/\delta_\mathrm{g}$ and holds for $\delta_\mathrm{g}\ll \nu_\mathrm{g}$. The cyan curve in Fig. \ref{frequencymixing}(b) corresponds to a vertical cut at $\nu_\mathrm{THz}/\delta_\mathrm{g}=1$. It can be derived from the figure, that for sampling of THz pulses centered at $\nu_\mathrm{THz}=1.3\times \delta_\mathrm{g}$, the SNR is expected to improve by a factor of 5.1 if the gate spectrum is high-pass filtered at $\nu_\mathrm{HP} = \nu_\mathrm{g}+\delta_\mathrm{g}$. 

\par

In conclusion, a universal method is introduced to enhance the sensitivity of EOS. It is shown, that our approach lowers the shot noise significantly and is able to additionally suppress technical noise below a reduced shot noise level. Existing EOS setups can be easily upgraded to enhance the SNR by up to 5 times saving as much as a factor 25 in measurement time, by adding a cost-effective spectral filter that can be selected with the presented analysis. Our idea will be instrumental for the emerging field of field-sensitive quantum optics where electro-optic sampling may ultimately be sophisticated to identify the absolute phase and amplitude of a few-photon state directly in the time domain.

Support by the German Research Foundation (DFG) via the Emmy Noether Program (HU1598/1-1) and the European Research Council via ERC Starting Grant QUANTUMsubCYCLE is acknowledged.


\end{document}